\documentclass[12pt]{article}
\usepackage{amsmath}
\usepackage{hyperref}
\usepackage{graphicx}
\usepackage{epsfig}
\usepackage{subfigure}
\usepackage{enumerate}
\usepackage{natbib}
\usepackage{listings}
\usepackage{float}
\usepackage{cases}
\usepackage{setspace}
\usepackage{url} 

\newcommand{\blind}{0}

\begin{document}

\def\spacingset#1{\renewcommand{\baselinestretch}%
{#1}\small\normalsize} \spacingset{1}

\doublespacing


\if0\blind
{
  \title{\bf Bayesian Analysis of Modified Weibull distribution under progressively censored competing risk model}
  \author{Arabin Kumar Dey\\
    Department of Mathematics, IIT Guwahati\\
    Abhilash Jha \\
    Department of Mathematics, IIT Guwahati\\
    Sanku Dey\\
    St. Anthony's College\\ 
}
  \maketitle
} \fi

\if1\blind
{
  \bigskip
  \bigskip
  \bigskip
  \begin{center}
    {\LARGE\bf Bayesian Analysis of Modified Weibull distribution under progressively censored competing risk model}
\end{center}
  \medskip
} \fi

\bigskip
\begin{abstract}
 In this paper we study bayesian analysis of Modified Weibull distribution under progressively censored competing risk model.  This study is made for progressively censored data.  We use deterministic scan Gibbs sampling combined with slice sampling to generate from the posterior distribution.  Posterior distribution is formed by taking prior distribution as reference prior.  A real life data analysis is shown for illustrative purpose.   
\end{abstract}

\noindent%
\vfill

\newpage
\spacingset{1.45} 
\section{Introduction}
\label{sec:intro}

  In survival and medical studies it is quite common that more than one cause of failure may be directed to a system at the same time.  It is often interesting that an investigator needs to estimate a specific risk in presence of other risk factors.  In statistical literature analysis of such risk is known as competing risk model.  Parametric inference of competing risk models are studied by many authors assuming that competing risks follow different lifetime distributions such as gamma, exponential and Weibull distribution; see for example \cite{BerksonElveback:60}, \cite{cox:59}, \cite{DavidMoeschberger:78}.  However determination of the cause of the failure is more difficult many times to observe than to follow up the time to failure.  Under the assumption that risks follow exponential distribution, inference of the model studied by \cite{miyakawa:84}.  \cite{kundubasu:00} considered the same model, studied by Miyakawa under the assumption that every member of a certain target population either dies of a particular cause, say cancer or by other causes. Model can be more flexible by considering the fact that some individual may be alive at the end of experiment i.e. data are censored.  Such models are available in literature even under bayesian set up; see for example \cite{KunduPradhan:11}, \cite{PareekKunduKumar:09}. 

 Here we consider competing risk under progressive type-II censoring.  The censoring scheme is defined as follows.  Consider n individuals in a study and assume that there are k causes of failure which are known.  At time of each failure, one of more surviving units may be removed from the study at random.  The data from a progressively type-II censored sample is as follows :  $ (X_{1:m:n}, \delta_{1}, R_{1}), \cdots, (X_{m:m:n}, \delta_{m}, R_{m}). $ $X_{1:m:n} < \cdots < X_{m:m:n}$.  Note that the complete and type-II right censored samples are special cases of the above scheme when $R_{1} = \cdots = R_{m} = 0$ and $R_{1} = R_{2} = \cdots = R_{m - 1} = 0$ and $R_{m} = n - m$ respectively.  For an exhaustive list of references and further details on progressive censoring, the reader may refer to the book by \cite{BalakrishnanAggarwala:2000}.  

  The main focus of this paper is the analysis of parametric competing risk model when the data is progressively censored.  No work has been done taking modified Weibull as parametric distribution of cause of the failures.  Modified Weibull has different forms.  We consider a specific form mentioned in section 2 which is not well-studied.  In bayesian analysis we choose reference prior.  Use of reference prior is rare as it makes the computational problem harder, though the best motivation for prior selection can be obtained through reference prior. Also many times expression of the prior may not be tractable.  However using gibbs combined with slice can provide a solution to this problem. Bayes estimators and credible intervals are also obtained.  

 The organization of the chapter is as follows.  In section 2 we describe the model and present the definitions and notation used throughout the paper.  The bayesian estimation of the different parameters are considered in Section 3.  Numerical results are provided in section 4.  We illustrate the performance of those techniques in section 5 using real data set.  Finally some conclusions are drawn in Section 6.

\section{Section 2}
\label{sec:model}

 We assume $(X_{1i}, X_{2i}),  i = 1, \cdots, n$ are n independent, identically distributed (i.i.d.) modified Weibull random variable.  Further $X_{1i}$ and $X_{2i}$ are independent for all $i = 1, \cdots, n$ and $X_{i} = \min\{ X_{1i}, X_{2i} \}$.  We observe the sample $ (X_{1:m:n}, \delta_{1}, R_{1}), \cdots, (X_{m:m:n}, \delta_{m}, R_{m}) $ assuming all the causes of failure are known.  We assume that the $X_{ji}$'s are modified Weibull distribution with parameters $\lambda_{j}$ for $i = 1, \cdots, n$ and for $j = 1, 2$.  The distribution function $F_{j}(\cdot)$ of $X_{ji}$ has the following form : $ F_{j}(t) = 1 - e^{\lambda_{j} \alpha(1 - e^{(\frac{t}{\alpha})^{\beta}})}$, 
$j = 1$ and 2.  
Therefore for each cause $j$, the pdf of failure time can be given by,
 $ f_{j}(t) = \lambda \beta (\frac{t}{\alpha})^{(\beta - 1)}e^{(\frac{t}{\alpha})^{\beta}}e^{\lambda_{i}\alpha(1 - e^{(\frac{t}{\alpha})^{\beta}})} $  
  
 The likelihood function of the observed data is 
\begin{eqnarray*}
L(\lambda_{1}, \lambda_{2}, \beta, \alpha) & = & \prod_{i = 1}^{m} S^{R_{i}}_{T}(t_{i}; \alpha, \beta, \lambda_{1} + \lambda_{2}) \prod_{i = 1}^{m_{1}} [f_{1}(t_{i}; \alpha, \beta, \lambda_{1})S_{2}(t_{i}; \alpha, \beta, \lambda_{2})]\\ & \times & \prod_{i = m_{1}}^{m} [f_{2}(t_{i}; \alpha, \beta, \lambda_{1})S_{1}(t_{i}; \alpha, \beta, \lambda_{2})]
\end{eqnarray*}

where $S_{T}(t_{i}; \alpha, \beta, \lambda_{1} + \lambda_{2}) = (1 - F_{T}(t_{i}; \alpha, \beta, \lambda_{1} + \lambda_{2})).$

 Evaluating the above likelihood we get 
\begin{eqnarray*}
L(\lambda_{1}, \lambda_{2}, \beta, \alpha) & = & n(n - R_{1} - 1)\cdots(n - R_{1} - \cdots - R_{m-1} - m + 1)\\ &\times & \lambda^{m_{1}}_{1}\lambda^{m - m_{1}}_{2} \beta^{m} e^{\sum_{i = 1}^{m} (\frac{t_{i}}{\alpha})^{\beta}}[\prod_{i = 1}^{m} (\frac{t_{i}}{\alpha})^{\beta - 1}] e^{(\lambda_{1} + \lambda_{2})\alpha\sum_{i = 1}^{m}(R_{i} + 1)(1 - e^{(\frac{t_{i}}{\alpha})^{\beta}})} 
\end{eqnarray*}

\subsection{Prior Assumption :}
\label{sec:prior}
 
 The key idea is to derive reference prior described by \cite{BergerJamesBernardo:92} (1992) is to get prior $\pi(\theta)$ that maximizes the expected posterior information about the parameters.  Let $I(\theta)$ is expected information about $\theta$ given the data $X = \underline{x}$.  Then
$$ I(\theta) = E_{X}(K(p(\theta | \underline{x}) || \pi(\theta))) $$  where $K(p(\theta | \tilde{x}) || \pi(\theta))$ is the Kullback-Leibler distance between posterior $p(\theta| \underline{x})$ and prior $p(\theta)$ can be given by $$ K(\pi(\theta | \underline{x}) || \pi(\theta)) = \int_{\Omega_{\theta}}^{} \ln \frac{\pi(\theta | \underline{x})}{\pi(\theta)} \pi(\theta | \underline{x}) d\theta $$

  If the set up is as follows, $\theta = (\theta_{1}, \theta_{2})$, where $\theta_{1}$ is $p_{1} \times 1$ and $\theta_{2}$ is $p_{2} \times 1$.  We define $p = p_{1} + p_{2}.$ where $\theta_{1}$ is the parameter of interest and $\theta_{2}$ is a nuisance parameter. 
  
 Let $I(\theta) = I(\theta) = \left(\begin{array}{cc} I_{11}(\theta) & I_{12}(\theta)\\ I_{21}(\theta) & I_{22}(\theta)   \end{array} \right)$

 We can show $\pi(\theta_{2} | \theta_{1}) = |I_{22}(\theta)|^{\frac{1}{2}} c(\theta_{1})$, where $c(\theta_{1})$ is the constant that makes this distribution a proper density.  We will be using this fact to get the expression for the prior in conditional distribution at each Gibbs sampler step.  

\noindent{\textbf{Reference Prior for Gibbs sampling Set-up}}

  In the Gibbs sampling set-up we try to reduce the posterior as the conditional distribution of one parameter given the other parameters and the data.  For example, we are interested in finding the $\displaystyle p(\lambda_{1} | \lambda_{2}, \beta, \alpha, x) \propto p(x | \lambda_{1}, \lambda_{2}, \beta, \alpha) p_{\text{reference prior}}(\lambda_{1} | \lambda_{2}, \beta, \alpha) .$  Complication of finding reference prior can be much simplied by using Gibbs sampler method.  

\begin{eqnarray*}
\pi(\lambda_{1} | \lambda_{2}, \beta, \alpha) & \propto & \sqrt{\mathcal{J}(\lambda_{1} | \lambda_{2}, \beta, \alpha, x)}\\
           & = & \sqrt{-\left(\frac{\partial^{2} l}{\partial \lambda^{2}_{1}}\right)}\\
\end{eqnarray*}
\begin{eqnarray*}
\pi(\beta | \lambda_{1}, \lambda_{2}, \alpha) & \propto & \sqrt{\mathcal{J}(\beta| \lambda_{1}, \lambda_{2}, \alpha, x)}\\
           & = & \sqrt{-\left(\frac{\partial^{2} l}{\partial \beta^{2}}\right)}
\end{eqnarray*}
\begin{eqnarray*}
\pi(\alpha | \beta, \lambda_{1}, \lambda_{2}) & \propto & \sqrt{\mathcal{J}(\alpha| \beta, \lambda_{1}, \lambda_{2}, x)}\\
           & = & \sqrt{-\left(\frac{\partial^{2} l}{\partial \alpha^{2}}\right)}\\
\end{eqnarray*}
\begin{eqnarray*}
\pi(\lambda_{2} | \beta, \lambda_{1}, \alpha) & \propto & \sqrt{\mathcal{J}(\lambda_{2} | \beta, \lambda_{1}, \alpha, x)}\\
           & = & \sqrt{-\left(\frac{\partial^{2} l}{\partial \lambda^{2}_{2}}\right)}\\
\end{eqnarray*}

\section{Posterior Analysis and Bayes Inference}
\label{sec:posterior}

 However generation of random number from these conditional densities is not tractable through inverse transform method due to its highly complicated form.  Therefore we use Slice sampler to generate sample from those conditional distribution.  Steps of the algorithms and calculation of HPD region can be provided as follows :

\subsection{Algorithm}
\begin{enumerate}
\item Choose a starting value of $\lambda_1, \lambda_2, \alpha, \beta$.
\item Use slice sampling to generate $\lambda_1$ using above $\lambda_2, \alpha, \beta$.
\item Use this new $\lambda_1$ and earlier $\lambda_2, \alpha, \beta$ to generate new $\lambda_2$.
\item Use new $\lambda_1, \lambda_2$ and earlier $\alpha, \beta$ to generate new $\alpha$.
\item Use new $\lambda_1, \lambda_2, \alpha$ to generate new $\beta$.
\item repeat step 2-5 M times to generate new ($\lambda_{1i}, \lambda_{2i}, \alpha_i, \beta_i$) $i = 1, \cdots, m$.
\item Mean Bayesian estimate of $\lambda_1, \lambda_2, \alpha, \beta$ is given by:-
$$ \hat{\lambda}_{1B} = \frac{1}{M}\sum_{k=1}^M \lambda_{1k}, \hat{\lambda}_{2B} = \frac{1}{M}\sum_{k=1}^M \lambda_{2k} $$
$$ \hat{\alpha}_{B} = \frac{1}{M}\sum_{k=1}^M \alpha_{k}, \hat{\beta}_{B} = \frac{1}{M}\sum_{k=1}^M \beta_{k} $$
\item To obtain HPD(Highest Posterior Density) region of $\lambda_1$, we order $\{\lambda_{1i}\}$, as $\lambda_{1(1)} < \lambda_{1(2)} < \cdots  < \lambda_{1(M)}$. Then 100(1 - $\gamma$)$\%$ HPD region of $\lambda_1$ become
$$(\lambda_{1(j)}, \lambda_{1(j + M - M\gamma)}), \qquad for \quad j = 1, \cdots , M\gamma$$

Therefore 100(1 - $\gamma$)$\%$ HPD region of $\lambda_1$ becomes $(\lambda_{1(j^*)},\lambda_{1(j^* + M - M\gamma)}),$ where $j^*$ is such that $$\lambda_{1(j^* + M - M\gamma)} - \lambda_{1(j^*)} \leq \lambda_{1(j + M - M\gamma)} - \lambda_{1(j)}$$
for all $j = 1, \cdots , M\gamma$. Similarly, we can obtain the HPD credible interval for $\lambda_2$, $\alpha$ and $\beta$.
\end{enumerate}

\section{Numerical Simulation:}
\label{sec:numsimu}

 In this section we conduct a simulation study to investigate the performance of the proposed Bayes estimators.  We generate sample points from the actual distribution assuming the parameters provided along with different censoring schemes and then apply it to our model to predict the parameters and compare the results.  To generate data sets currently we assign the cause of failure as 1 or 2 with probability 0.5.  We have generated 200 data points. Then we have divided it in 2 cases.  In case 1 we consider all 200 data points (which is \textbf{Type 2 Censoring}) and in case 2 we will apply censoring on those data (which is \textbf{Progressive censoring}) and then predict the parameters.  Below are the different schemes of data sets used for validation.

\begin{enumerate}[{\textbf{Scheme}} 1]
  \item $\lambda_1=1, \lambda_2=0.6, \alpha=0.3, \beta=0.1, n=200$, Type 2 Censoring
  \item $\lambda_1=1, \lambda_2=0.6, \alpha=0.3, \beta=0.1, n=200$, Progressive Type 2 Censoring
  \item $\lambda_1=1, \lambda_2=1, \alpha=0.3, \beta=0.1, n=200$, Type 2 Censoring
  \item $\lambda_1=1, \lambda_2=1, \alpha=0.3, \beta=0.1, n=200$, Progressive Type 2 Censoring
\end{enumerate}

\noindent{\textbf{Scheme 1}}

\begin{figure}[H]
    \centering
    \includegraphics[width=15cm]{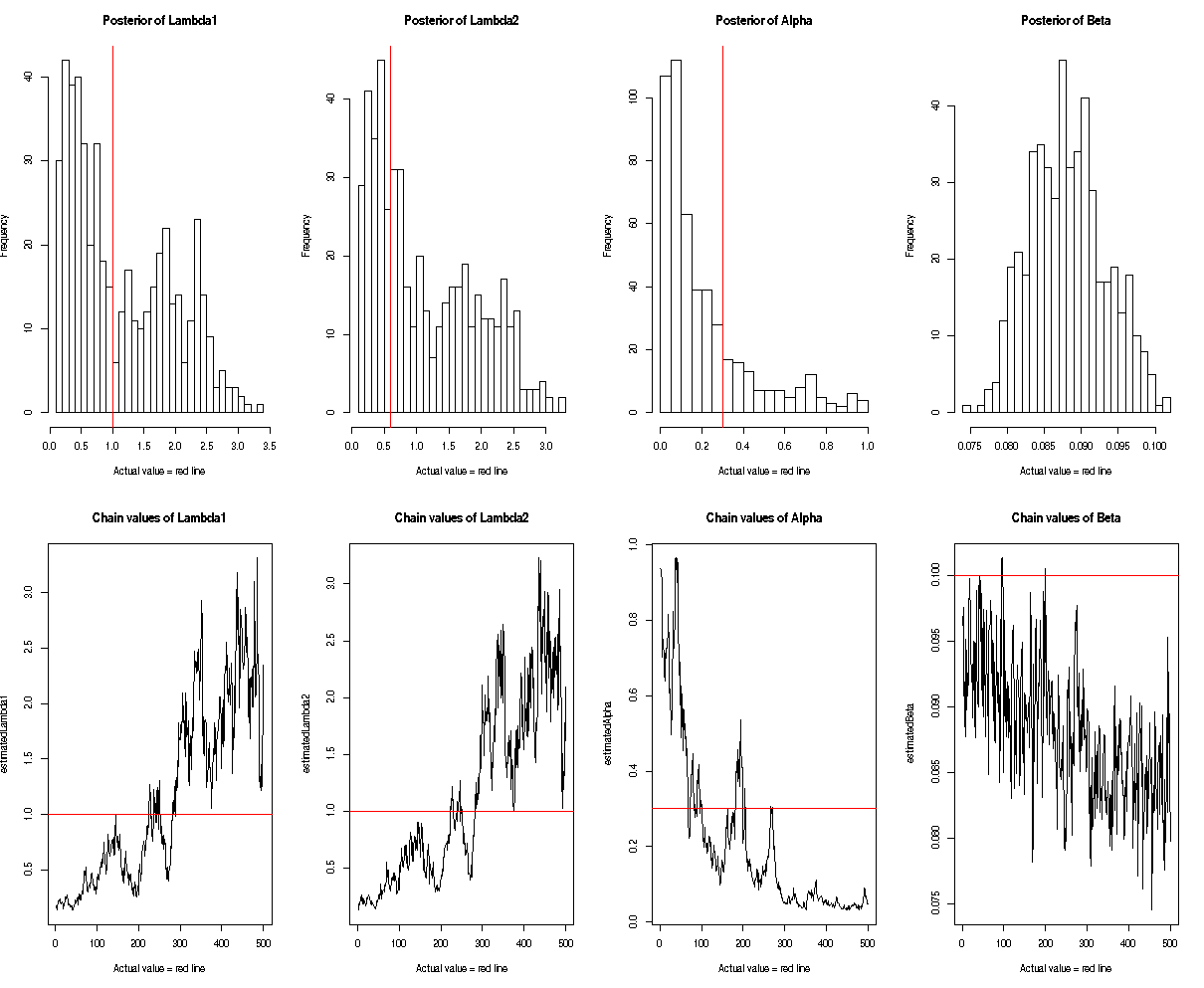}
    \caption{Histogram and trace plot of parameter values for Scheme 1}
\end{figure}

\begin{table}[H]
\centering
\begin{tabular}{ |p{3cm}|p{3cm}|p{4cm}|  }
 \hline
 Parameters & Actual & Result \\
 \hline
 $\lambda_1$ &	1	&	0.88139633304646\\
 $\lambda_2$ &	0.6	&	0.87962618837108\\
 $\alpha$    &	0.3	&	0.226456918989323\\
 $\beta$     &	0.1	&	0.087896231075444\\
 \hline
\end{tabular}
\caption{Results Comparison for Scheme 1}
\end{table}

\noindent{\textbf{Scheme 2}}

\begin{figure}[H]
    \centering
    \includegraphics[width=15cm]{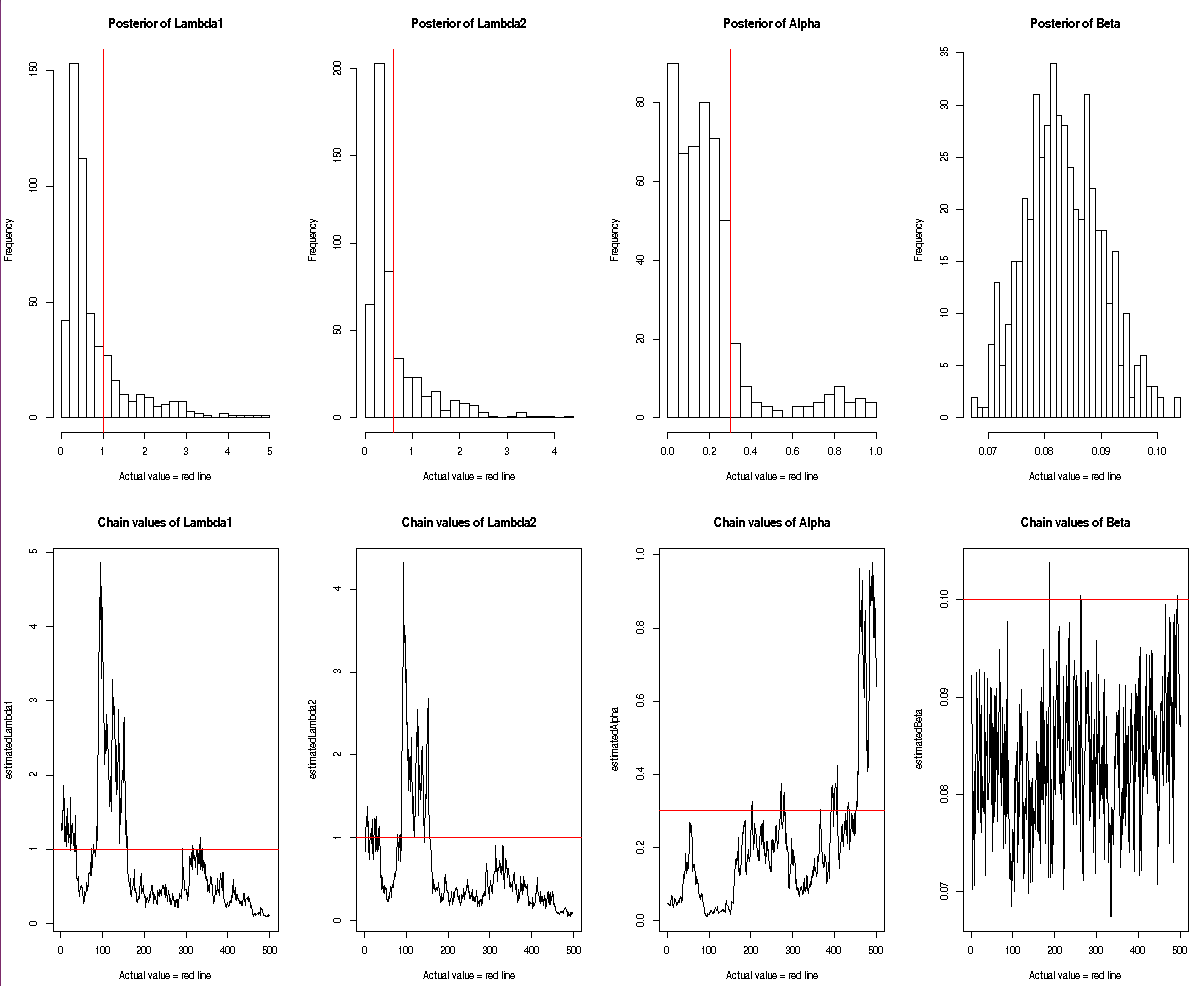}
    \caption{Histogram and trace plot of parameter values for Scheme 2 }
\end{figure}

\begin{table}[H]
\centering
\begin{tabular}{|p{3cm}|p{3cm}|p{3cm}|}
 \hline
 Parameters & Actual & Result \\
 \hline
$\lambda_1$ &		1	&	0.7918084\\
$\lambda_2$   &		0.6	&	0.6261012\\
$\alpha$&		0.3	&	0.2062639\\
$\beta$	&		0.1	&	0.08331048\\
 \hline
\end{tabular}
\caption{Results Comparison for Scheme 2}
\end{table}


\noindent{\textbf{Scheme 3}}

\begin{figure}[H]
    \centering
    \includegraphics[width=15cm]{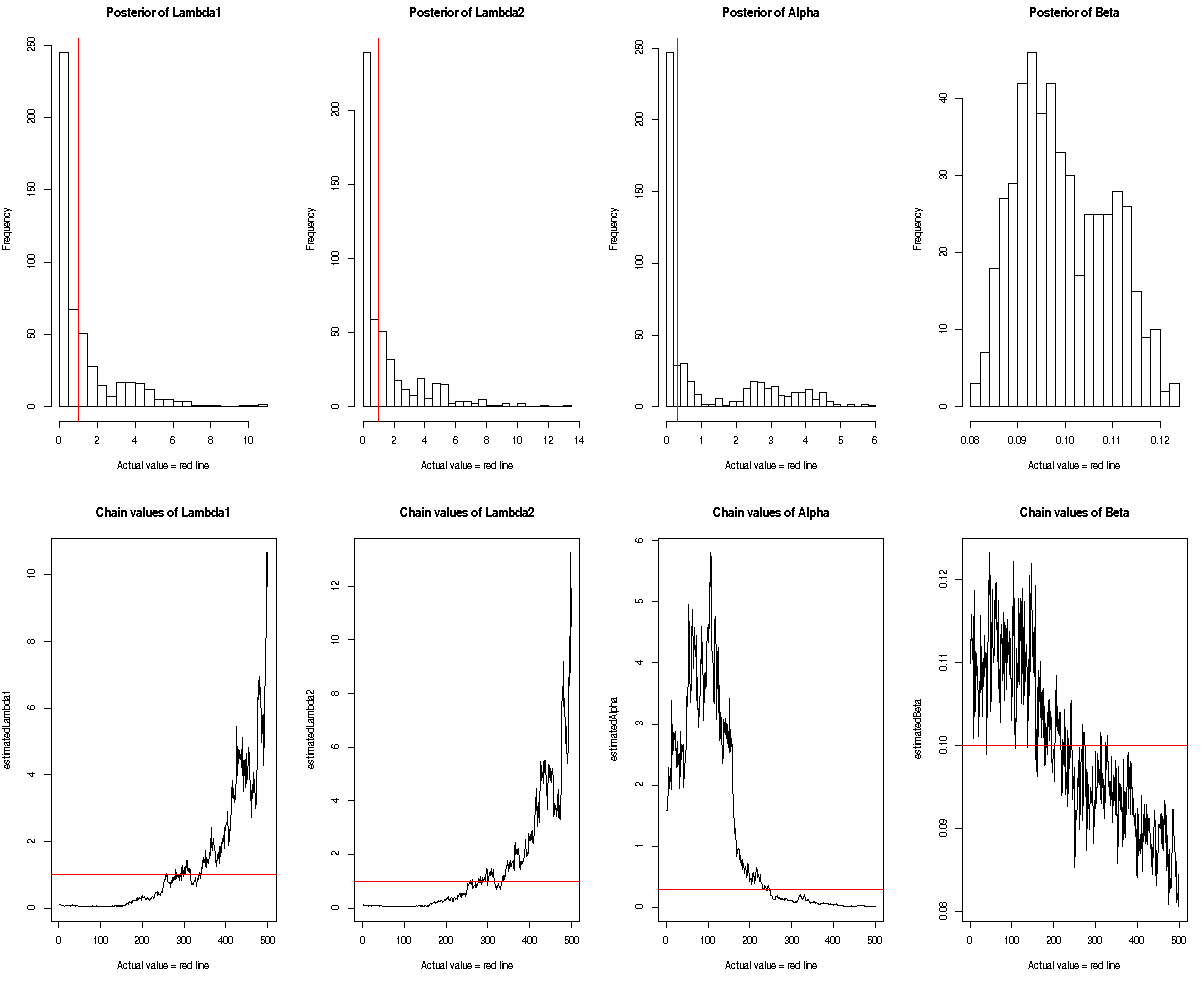}
    \caption{Histogram and trace plot of parameter values for Scheme 3}
\end{figure}

\begin{table}[H]
\centering
\begin{tabular}{|p{3cm}|p{3cm}|p{4cm}|}
 \hline
 Parameters& Actual &Result \\
 \hline
 $\lambda_1$	&	1		&	0.6300016260981249\\
 $\lambda_2$	&	1	    	&	0.698699693413326\\
 $\alpha$	&	0.3		&	0.215528915817512\\
 $\beta$	&	0.1		&	0.0977690735863656\\
 \hline
\end{tabular}
\caption{Results Comparison for Scheme 3}
\end{table}

\noindent{\textbf{Scheme 4}}

\begin{figure}[H]
    \centering
    \includegraphics[width=15cm]{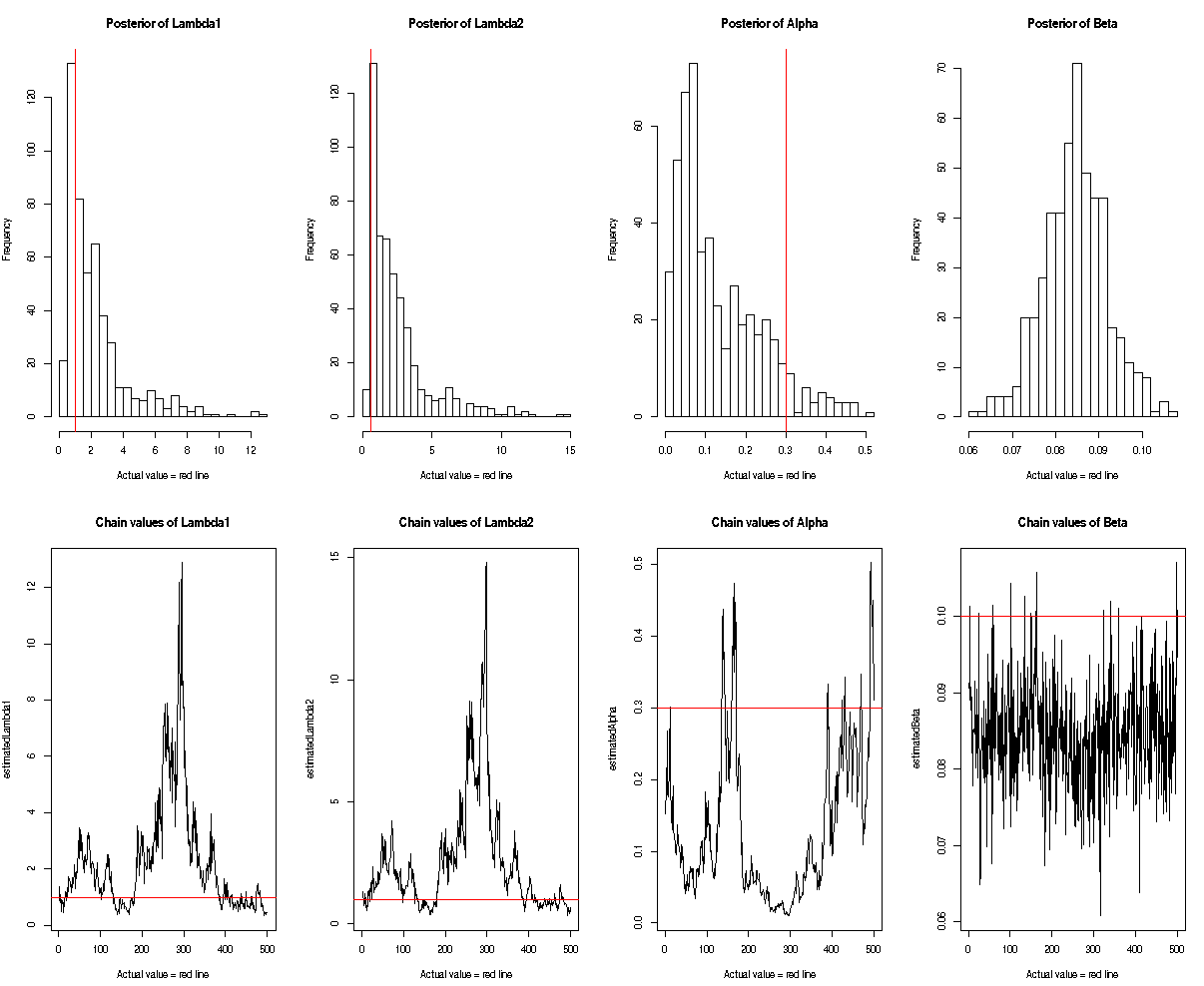}
    \caption{Histogram and trace plot of parameter values for Scheme 4}
\end{figure}

\begin{table}[H]
\centering
\begin{tabular}{|p{3cm}|p{3cm}|p{3cm}|}
 \hline
 Parameters  & Actual &Result \\
 \hline
 $\lambda_1$ &   1	&   0.918084\\
 $\lambda_2$  &   1	&   0.8261012\\
 $\alpha$  &   0.3	&   0.3062639\\
 $\beta$ &   0.1	&   0.08331048\\
 \hline
\end{tabular}
\caption{Results Comparison for Scheme 4}
\end{table}
 
\section{Data Analysis : Application on Follicular Cell Lymphoma Data}
\label{sec:data}

  Now We will move on to apply our model on real life data. For that we have considered the follicular cell lymphoma data from Pintilie (2007) where additional details about data set can be found. The data set can be downloaded from \href{http://www.uhnres.utoronto.ca/labs/hill/datasets/Pintilie/datasets/follic.txt}{follic.txt}, and consists of 541 patients with
early disease stage follicular cell lymphoma (I or II) and treated with radiation only (chemo = 0) or a combined treatment with radiation and chemotherapy (chemo = 1). Parameters recorded were path1, ldh, clinstg, blktxcat, relsite, chrt, survtime, stat, dftime, dfcens, resp and stnum. The two competing risks are \textbf{death without relapse} and \textbf{no treatment response}. The patient's ages (age: mean = 57 and sd = 14) and haemoglobin levels (hgb: mean = 138 and sd = 15) were also recorded. The median follow-up time was 5.5 years. There are more parameters which are not of our concern.

  First we read the data, compute the cause of failure indicator. Below is the code to calculate the cause. 

\begin{lstlisting}[language=R]
R> evcens <- as.numeric(fol$resp == "NR" | fol$relsite != "")
R> crcens <- as.numeric(fol$resp == "CR" & 
fol$relsite == "" & fol$stat == 1)
R> cause <- ifelse(evcens == 1, 1, ifelse(crcens == 1, 2, 0)
R> table(cause)
cause
0   1   2
193 272 76
\end{lstlisting}

  There are 272 (no treatment response or relapse) events due to the disease, 76 competing risk
events (death without relapse) and 193 censored individuals. The event times are denoted as dftime. Thus our data set is prepared.

  Now we will consider this data set and apply it on our model.  We will consider 3 cases of analysis as defined below :-
\begin{enumerate}[{Case} 1]
 \item  We will only consider those data in which cause = 1 and cause = 2 are considered. 
 \item  Here we will consider the censor data too.  Now to prepare data set we will first sort data according to time. Then we will remove data points between two non-zero causes. 
\end{enumerate}

  For Case 1, the following are the bayes estimate corresponding to the parameters, $\lambda_1$ = 0.12128, $\lambda_2$ = 0.01762, $\alpha$ = 2.33876, $\beta$ = 0.44871.  For Case 2, bayes estimates are, $\lambda_1$ = 0.034, $\lambda_2$ = 0.004, $\alpha$ = 30.228, $\beta$ = 0.536.

\begin{figure}[H]
    \centering
    \includegraphics[width=15cm]{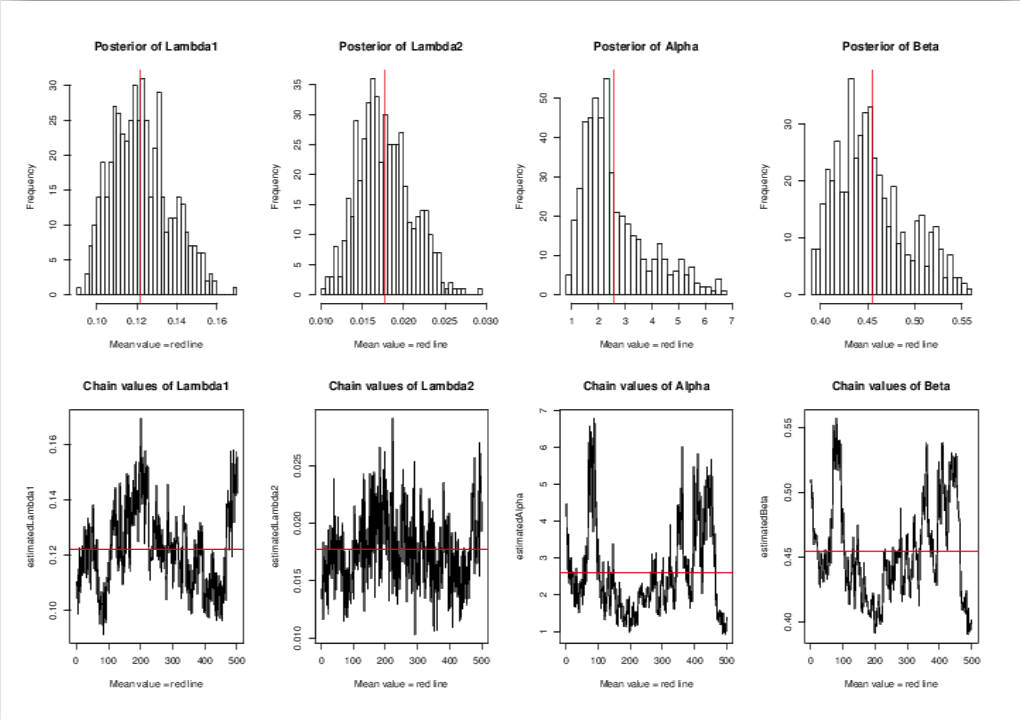}
    \caption{Histogram and trace plot of parameters for Follicular Data Case 1 }
\label{fig:case1}
\end{figure}

  We show the histogram and trace plot of the parameters of the Follicular Data Case 1 and Case 2 in \ref{fig:case1}, \ref{fig:case2} respectively. 

\begin{figure}[H]
    \centering
    \includegraphics[width=15cm]{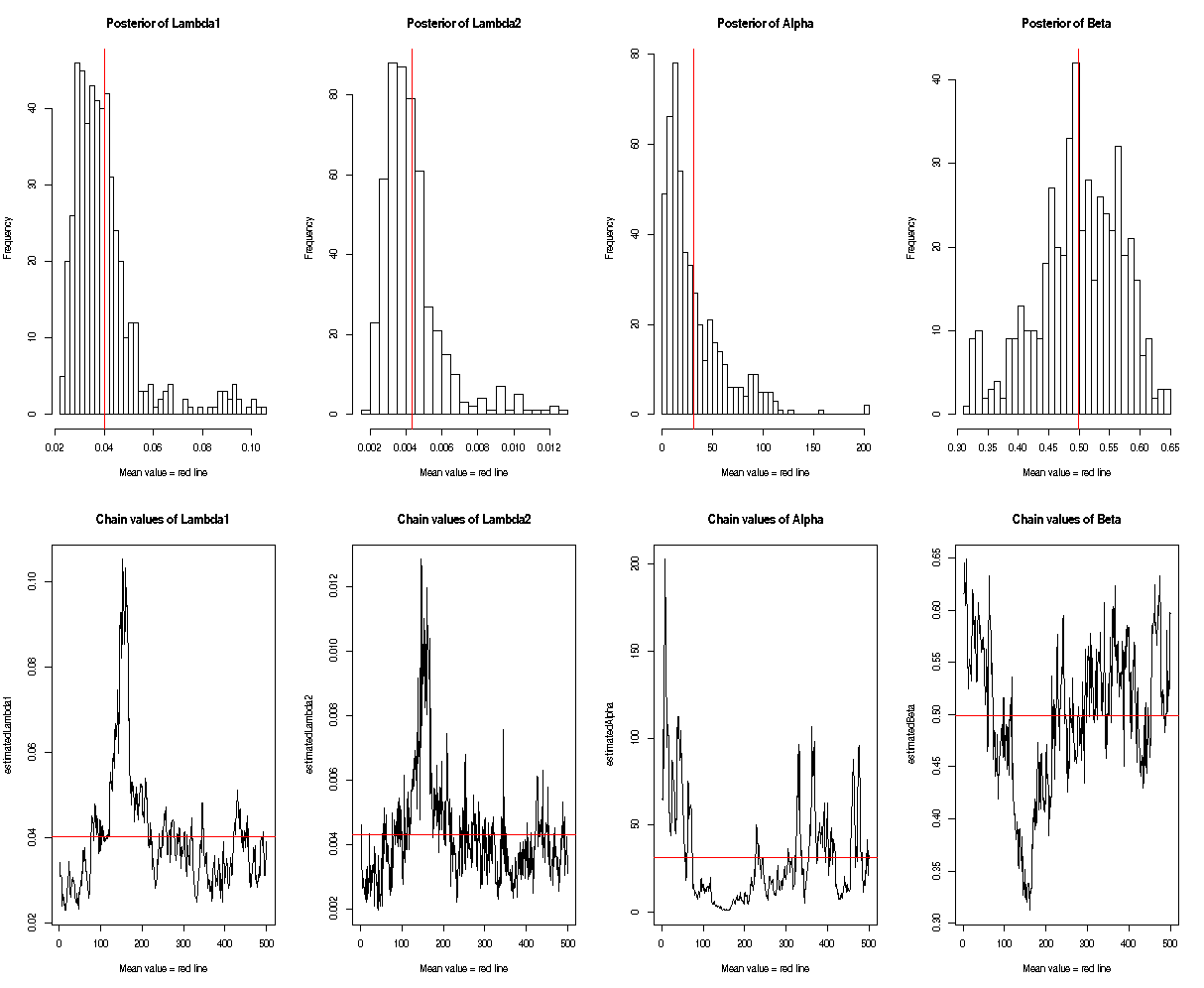}
    \caption{Histogram and trace plot of parameters for Follicular Data Case 2 }
\label{fig:case2}
\end{figure}

\begin{table}[H]
\centering
\begin{tabular}{ |p{3cm}|p{4cm}|p{3cm}|p{3cm}|  }
 \hline
 Parameters & posterior median & Lower HPD & Upper HPD \\
 \hline
 $\lambda_1$ & 0.034	&  0.0222 & 0.0432\\
 $\lambda_2$ & 0.004	&  0.00289 & 0.0058\\
 $\alpha$    & 30.228   & 7.725 & 164.4087 \\
 $\beta$     & 0.536	&  0.448 & 0.6362\\
 \hline
\end{tabular}
\label{fig:case2}
\caption{Posterior median and lower and upper HPD for Follicular Data in Case 2}
\end{table}

 The results obtained for HPD region of $\alpha$ apparently seem to look awkward as its width is very high.  However taking posterior median as true value of the parameters of modified Weibull, we can easily cross-check it provides a good fit for histograms of the distribution of cause-specific data.  This verifies the distributional assumption of this parametric approach.

\section{Conclusion}
\label{sec:conc}

  We consider the bayesian analysis of the competing risks data when they are Type-II progressively censored.  Numerical simulation shows the posterior median calculated via slice sampling combined with gibbs sampler works quite well. In this article we consider with two causes of failure only, the work can be extended to more than two causes of failure.  We select parameters to follow reference prior which is very flexible.  The similar work can be done for type-I progressively censored data.  The work is on progress.  Slice sampling uses step out methods where selection of width plays an important role.  An automated choice of such selection based on the data and distribution can enhance the quality of the algorithm.  

\bigskip
\begin{center}
{\large\bf SUPPLEMENTARY MATERIAL}
\end{center}


{}

\end{document}